\documentclass[prl,superscriptaddress,showpacs,reprint]{revtex4-2}
\usepackage[]{fontenc}
\usepackage[latin9]{inputenc}
\usepackage{textcomp}
\usepackage{amsmath}
\usepackage{graphicx}
\usepackage{amssymb}
\usepackage{bbm}
\usepackage{dsfont}
\usepackage{newtxmath}
\usepackage{dcolumn}
\usepackage{babel}
\usepackage{color}
\usepackage[normalem]{ulem}
\usepackage{pifont}
\usepackage{rotating}
\usepackage{float} %added by shivesh

\usepackage{hyperref}
\newcommand{\mpp}{MnPt$_{x}$Pd$_{1-x}$}
\newcommand{\la}{L1$_{0}$}

\definecolor{dark-red}{rgb}{0.9,0.15,0.15}
\definecolor{dark-blue}{rgb}{0.15,0.15,0.4}
\definecolor{medium-blue}{rgb}{0,0,0.5}
\hypersetup{
	colorlinks, linkcolor={black},
	citecolor={dark-blue}, urlcolor={medium-blue}
}

\newcommand{\I}{\mathbb{I}}

\begin{document}
	
\title{Four-fold Anisotropic Magnetoresistance in Antiferromagnetic Epitaxial Thin Films of MnPt$_{x}$Pd$_{1-x}$ }

\author{Shivesh Yadav}

\affiliation{Department of Condensed Matter Physics and Material Science, Tata Institute of Fundamental Research, Mumbai, MH 400005, India}

\author{Shikhar Kumar Gupta}

\affiliation{Department of Condensed Matter Physics and Material Science, Tata Institute of Fundamental Research, Mumbai, MH 400005, India}

\author{Mohit Verma}

\affiliation{School of Physical Sciences, Indian Institute of Technology Mandi, HP 175075, India}

\author{Debjoty Paul}

\affiliation{Department of Condensed Matter Physics and Material Science, Tata Institute of Fundamental Research, Mumbai, MH 400005, India}

\author{Abira Rashid}

\affiliation{Department of Materials Engineering, Indian Institute of
Technology, Gandhinagar, GJ 382055, India}

\author{Bhagyashree Chalke}

\affiliation{Department of Condensed Matter Physics and Material Science, Tata Institute of Fundamental Research, Mumbai, MH 400005, India}

\author{Rudheer Bapat}

\affiliation{Department of Condensed Matter Physics and Material Science, Tata Institute of Fundamental Research, Mumbai, MH 400005, India}

\author{Nilesh Kulkarni}

\affiliation{Department of Condensed Matter Physics and Material Science, Tata Institute of Fundamental Research, Mumbai, MH 400005, India}

\author{Abhay Gautam}

\affiliation{Department of Materials Engineering, Indian Institute of
Technology, Gandhinagar, GJ 382055, India}

\author{Arti Kashyap}

\affiliation{School of Physical Sciences, Indian Institute of Technology Mandi, HP 175075, India}

\author{Shouvik Chatterjee}

\email[Author to whom correspondence should be addressed: ]{shouvik.chatterjee@tifr.res.in}

\affiliation{Department of Condensed Matter Physics and Material Science, Tata Institute of Fundamental Research, Mumbai, MH 400005, India}
                                         
\begin{abstract}

Antiferromagnets are emerging as promising alternatives to ferromagnets in spintronics applications. A key feature of antiferromagnets is their anisotropic magnetoresistance (AMR), which has the potential to serve as a sensitive marker for the antiferromagnetic order parameter. However, the underlying origins of this behavior remains poorly understood, particularly, in thin film geometries. In this study, we report the observation of AMR in epitaxial thin films of the collinear \la\/ antiferromagnet \mpp\/. In the thicker films, AMR is dominated by a non-crystalline two-fold component, which emerges from domain reconfiguration and spin canting under applied magnetic field. As the film thickness is reduced, however, a crystalline four-fold component emerges, accompanied by the appearance of uncompensated magnetic moment, which strongly modifies the magnetotransport properties in the thinner films. We demonstrate that interfacial interactions lead to a large density of states (DOS) at the Fermi level. This enhanced DOS, combined with disorder in the thinner films, stabilizes the uncompensated moment and results in a four-fold modulation of the DOS as the Neel vector rotates, explaining the observed AMR behavior.
 
\end{abstract}

\maketitle

%%%%%%%%%%%%%%%%%%%%%%%%%%%%%%%%%%%%%%%%%%%%%%%%%%%%%%%%%%%%%%%%%%
%%%%Figure 1 %%%%%%%%%%%%%%%%%%%%%%%%%%%%%%%%%%%%%%%%%%%%%%%%%%%%%%%%%%%%%%%%%%

\begin{figure} [h!]
	\centering
	\includegraphics[width=1.0 \linewidth]{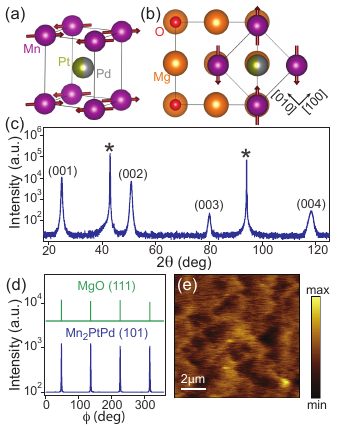}
	\caption{(a) Crystal structure of \mpp\/ (b) Epitaxial relationship of \mpp\/ with MgO substrate. (c) Out-of-plane $\theta$ - $2\theta$ scan of a 40 nm thick \mpp\/ thin film. Substrate peaks are marked by asterisks. (d) Azimuthal $\phi$-scan of the asymmetric planes of MgO(111) and \mpp\/(101) establishing the epitaxial relationship \mpp\/[110]$||$MgO[100]. (e) AFM image of a 10 nm thick sample with a measured RMS roughness of 1.2 nm over a 10$\mu m$$\times$10$\mu m$ field of view.}
	\label{fig:fig1}
\end{figure}

%%%%%%%%%%%%%%%%%%%%%%%%%%%%%%%%%%%%%%%%%%%%%%%%%%%%%%%%%%%%%%%%%%

\section{Introduction}

Anisotropic magnetoresistance (AMR) in ferromagnets have played an important role in our understanding of spin-dependent transport phenomena since its discovery more that a century ago\cite{thomson1857xix}. The angular dependence of resistivity as the magnetic field is rotated with respect to the current direction and specific crystallographic axis of single crystals, known as AMR, is brought about by a combination of factors including spin-orbit coupling, exchange fields, and crystal field splittings\cite{mcguire1975anisotropic,campbell1970spontaneous,kokado2012anisotropic,vyborny2009microscopic}. Since AMR is even in magnetization, it is also expected in antiferromagnets, which is only beginning to be explored. Recently, antiferromagnets have emerged as a potential active element in spintronics devices due to a number of advantages over conventional ferromagnetic materials including stability under external magnetic field, absence of stray field, and faster spin dynamics\cite{park2011spin,baltz2018antiferromagnetic,lopez2019picosecond,wadley2016electrical}. AMR in antiferromagnets can potentially be utilized to monitor its magnetic state and is therefore an attractive read-out scheme for antiferromagnetic memory devices\cite{wadley2016electrical,marti2014room,kriegner2016multiple,bodnar2018writing,siddiqui2020metallic,wang2020spin,gonzalez2024anisotropic,wang2019giant}.

Manganese containing metallic antiferromagnets such as MnPt and MnPd possess large magnetic moments due to manganese (Mn) 3$d$ shell and strong spin-orbit coupling due to heavy metals (platinum(Pt)/palladium(Pd)), and therefore, are expected to exhibit strong AMR response\cite{shick2010spin}. Recently, the AMR and Hall response in these compounds have been utilized to establish current and strain driven switching of the Neel vector\cite{yan2019piezoelectric,duttagupta2020spin,shi2020electrical}. However, the origin of AMR in thin films of these compounds remain poorly understood. Both MnPt and MnPd have a collinear antiferromagnetic ground state with a tetragonal \la\/ crystal structure (CuAu-I type, space group - $P4/mmm$)\cite{pal1968magnetic}. 
 It was recently predicted that by combining \la\/ structures of MnPt and MnPd it might be possible to realize an inverse Heusler structure (TiAl$_{3}$ type, space group - $I4/mmm$) Mn$_{2}$PtPd, where Pt and Pd are ordered\cite{sanvito2017accelerated}. Although, experimental realization of Mn$_{2}$PtPd with inverse Heusler structure  was claimed to be realized in the same work, adequate structural characterization data was not provided. A subsequent work on polycrystalline bulk samples found that Pt and Pd are disordered in Mn$_{2}$PtPd resulting in an \la\/ (CuAu-I type) crystal structure\cite{kumar2018crystal}. 

 Here, we have established \la\/ crystal structure in epitaxial thin films of Mn$_{2}$PtPd, referred henceforth as \mpp\/, synthesized on MgO(001) substrates. We show that AMR in these thin films exhibits a remarkable thickness dependence. In the thicker films, AMR is dominated by a two-fold symmetric non-crystalline component resulting from domain reconfiguration and spin canting on the application of magnetic field, whereas a strong four-fold crystalline component emerges in the thinner films. The thinner films also exhibit exchange bias even when the films are field-cooled from temperatures below the Neel temperature, indicating the presence of uncompensated moments (UM), which significantly impacts their magnetotransport properties. Our first-principles calculations reveal the crucial role of interfacial interactions between Mn atoms in \mpp\/ and oxygen(O) atoms in MgO substrate in bringing about these phenomena. It results in a large enhancement in the density of states(DOS) at the Fermi level, which also stabilizes uncompensated moment in the thinner films. Coherent rotation of the Neel vector driven by the presence of UM leads to a significant modulation of the density of states giving rise to a four-fold symmetric AMR, which is absent in the thicker films, in accordance with the experimental results.

%Structural Properties%%%%%%%
\section{Results and Discussion}

\subsection{A. Epitaxial Synthesis and Structural Characterization}

Epitaxial \mpp\/ thin films were synthesized on MgO(001) substrates using radio-frequency (RF) magnetron sputtering from a stoichiometric target of Mn$_{2}$PtPd. Samples were grown at 450\textdegree C using 50 Watt RF power at 5 mTorr argon partial pressure. They were subsequently cooled down to room temperature and capped with a protective AlO$_{x}$ layer\cite{chatterjee2021controlling,chatterjee2021identifying} and finally post-annealed at 800\textdegree C for three hours. Single-crystalline, epitaxial nature of our thin films is confirmed by x-ray diffraction measurements, shown in Fig.\ref{fig:fig1}. By undertaking a detailed structural analysis, described in the Supplementary Information\cite{suppl}, we confirmed that Pt and Pd are disordered in epitaxial \mpp\/ thin films, which crystallize in an \la\/ structure\cite{suppl}, contrary to what has been suggested in an earlier report\cite{sanvito2017accelerated}. Our measurements are, however, in accordance with an earlier report on polycrystalline bulk samples\cite{kumar2018crystal}. Accordingly, all Miller indices are specified with respect to the \la\/ structure with $P4/mmm$ space group symmetry unless noted otherwise. \mpp\/ thin films are rotated 45 degrees in-plane with respect to MgO(001) substrates with an epitaxial relationship \mpp\/[110](001)$\parallel$ MgO[100](001) (Fig.\ref{fig:fig1}(b)), established through the x-ray diffraction measurements shown in Fig.\ref{fig:fig1}(c,d). The in-plane (\textit{a}) and out-of-plane (\textit{c}) lattice parameters were found to be 2.89 $\pm$ 0.01 and 3.59 $\pm$ 0.01\AA, respectively, for all film thicknesses\cite{suppl}, which is similar to the reported values for polycrystalline bulk \mpp\/ samples ($a=2.86$\AA\/ and $c=3.62$\AA)\cite{kumar2018crystal}.

%Magnetic properties%%%%%%%%%%%%%%%
\subsection{B. Magnetic Properties of \mpp\/ Thin Films}

The antiferromagnetic ground state in \la\/ alloys of Mn, Pt, and Pd is well established\cite{pal1968magnetic,yan2019piezoelectric,duttagupta2020spin,shi2020electrical,andresen1965equiatomic,kjekshus1967equiatomic,kren1968magnetic,umetsu2002electrical,umetsu2003magnetic,umetsu2006electrical,hama2007spin,gu2023magnetotransport,umetsu2006magnetic,park2019strain,su2020voltage,kumar2018crystal}. Neutron diffraction measurements on bulk polycrystalline \la\/ \mpp\/ samples has revealed a c-type antiferromagnet as the ground state (see Figs.~1(a,b) and S8 in the Supplementary Information\cite{suppl}), where the Neel vector lies in the $a$-$b$ plane\cite{kumar2018crystal}. This is also confirmed by our density functional theory calculations\cite{suppl}. High temperature susceptibility measurements in \mpp\/ thin films confirm a Neel temperature($T_{N}$) of 840 K\cite{suppl}, which, as expected, is intermediate between the reported values of $T_{N}$ for MnPt(970 K) and MnPd(780 K)\cite{pal1968magnetic,kren1968magnetic,kjekshus1967equiatomic,andresen1965equiatomic}, and similar to the observation in bulk polycrystalline \mpp\/ samples\cite{kumar2018crystal}. 
 
\begin{figure} 
	\centering
 \includegraphics[width=0.50\textwidth]{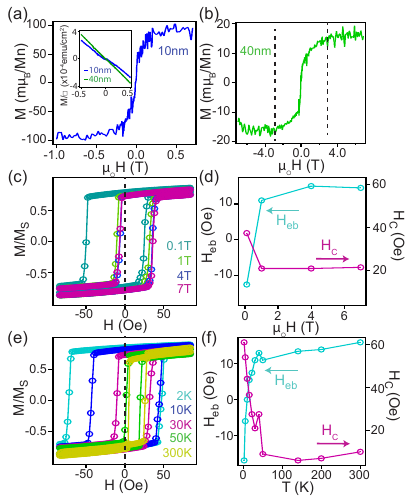}
	\caption{Magnetization as a function of magnetic field recorded at 100 K for (a) 10 and (b) 40 nm thick \mpp\/ films. Diamagnetic contribution of the MgO substrate has been subtracted. Inset in (a) shows magnetization per unit area for 10 and 40 nm thick films before subtraction of the diamagnetic contribution from the substrate. Magnetization hysteresis loop in \mpp\/(10 nm)/Fe(7 nm) bilayer sample measured (c) at 30 K under different cooling magnetic field and (e) at different temperatures under a cooling field of 1 T. Corresponding variation of the exchange ($H_{eb}$) and coercive ($H_{c}$) fields are shown in (d) and (f), respectively.}
	\label{fig:fig2}
\end{figure}

Magnetic field dependence of magnetization was measured at few different temperatures for 40 and 10 nm thick films. Representative data taken at 100 K is shown in Fig.~\ref{fig:fig2}(a-b). Spontaneous magnetization, which establish the presence of UM, is found only in the 10 nm thick films. The UM saturates to a value of 80 $m\mu_{B}$/Mn at a magnetic field of $\approx$ 2 KOe. Similar magnetization response is observed for magnetic field applied along different crystallographic directions\cite{suppl}. In contrast, only a dominant diamagnetic signal of the substrate is observed for the 40 nm thick films establishing nearly compensated antiferromagnetic order in the thicker films. After subtracting a linear background, we can identify a very weak magnetic susceptibility response, close to the detection limit. The observed weak, gradual increase in magnetic susceptibility as a function of magnetic field (see Fig.~2(b)), as opposed to a sharp change expected at a spin-flop transition, is attributed to the field induced change in the domain configuration, discussed in the next section. The minuscule step-like magnetization signal at zero field, is most likely related to magnetic impurity present in commercial substrates, which is frequently observed\cite{gonzalez2024anisotropic}.
 
 We also performed exchange bias measurements where bi-layers of \mpp\/(10/40nm)/Fe(7 nm) were cooled down from 380 K under a magnetic field along MgO(100)/\mpp\/(110), which is also the in-plane easy axis of the Fe layer.  We observe exchange bias effect only in the 10 nm thick films. The exchange bias field($H_{eb}$) is 11 Oe at 30 K, when the sample is field cooled under a magnetic field of 1 T. Under these conditions the estimated interfacial exchange coupling energy J$_{ex,int}$ is 0.047 erg/cm$^{2}$\cite{suppl}, which is similar to the recently reported value of 0.029 erg/cm$^{2}$ for Py/MnPd interface\cite{gu2023magnetotransport}, where similar evidence for uncompensated moment was found. However, this is much smaller than those of MnPd/Fe and MnPd$_{3}$/Co\cite{zhan2010plane,nam2007giant}. It is to be noted that the samples were cooled down from 380 K, a temperature much below the Neel temperature. Hence, the observed exchange bias is primarily driven by the presence of both pinned and  unpinned UM. This explains the absence of exchange bias in the 40 nm thick film and a relatively smaller value of $J_{ex,int}$ in the 10 nm thick films. The temperature and magnetic field dependence of exchange bias in the 10 nm thick films can be understood by considering the behavior of UM and its susceptibility to the applied field. At 30 K, $H_{eb}$ is negative when  magnetic field under which the sample is cooled is small, but becomes positive for cooling fields of 1 T and larger, as shown in Fig.~\ref{fig:fig2}(c,d). The observation of negative exchange bias at low cooling fields can be understood by assuming an antiferromagnetic spin alignment at the interface when the coupling energy of the AFM UM with the field is smaller than the AFM anisotropy energy. However, under a strong enough cooling field, the coupling energy of the AFM UM with the field can become stronger than the AFM anisotropy energy, which then orients along the magnetic field resulting in a ferromagnetic coupling at the interface, thereby frustrating the interaction at the AFM/FM interface. Under such a scenario, upon reversal of magnetic field direction a smaller field is required to reorient the FM spins along the field compared to reorienting them back to the initial direction. This results in a positive shift of the M-H loop, hence a positive $H_{eb}$\cite{nogues1996positive,arenholz2006magnetization}. Similarly, temperature dependence of AFM anisotropy can also lead to a sign reversal of $H_{eb}$ as a function of temperature, as observed under a 1 T cooling field, shown in Fig.~\ref{fig:fig2}(e,f). Here, with increasing temperature AFM anisotropy gets weaker, and at around T $\approx$ 30K, $H_{eb}$ changes from negative to positive. Therefore, squid magnetometry and exchange bias measurements unambiguously establish the presence of UM in the 10 nm thick films, which could not be detected in the 40 nm thick films. 

%%%%%%%%%%%%%%%%%%%%%%%%%%%%%%%%%%%%%%%%%%%%%%%%%%%%%%%%%%%%%%%%%%
%%%%Figure 2 %%%%%%%%%%%%%%%%%%%%%%%%%%%%%%%%%%%%%%%%%%%%%%%%%%%%%%%%%%%%%%%%%%

%%%%%%%%%%%%%%%%%%%%%%%%%%%%%%%%%%%%%%%%%%%%%%%%%%%%%%%%%%%%%%%%%%

%%%%%%%%%%%%%%%%%%%%%%%%%%%%%%%%%%%%%%%%%%%%%%%%%%%%%%%%%%%%%%%%%%
%%%%Figure 3 %%%%%%%%%%%%%%%%%%%%%%%%%%%%%%%%%%%%%%%%%%%%%%%%%%%%%%%%%%%%%%%%%%

\begin{figure} [h!]
	\centering
 \includegraphics[width=1\linewidth]{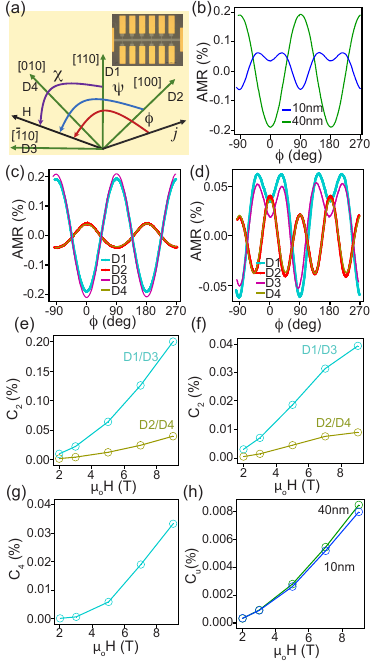}
	\caption{(a) Schematic of the orientation of different Hall bar devices used in this study. The relevant angles are also shown. The inset shows an optical micrograph image of a Hall bar device (b) AMR of 10 and 40 nm thick D1 devices oriented along \mpp\/[110] direction. (c-d) Comparison of AMR measured in D1(blue), D2(red), D3(magenta), and D4(green) devices for (c) 40 and (d) 10 nm thick films. Evolution of the two-fold symmetric component, $C_{2}$, with magnetic field in (e) 40nm (f) 10 nm thick films. (g) Magnetic field dependence of the four-fold symmetric component, $C_{4}$, in 10 nm thick films. (h) Evolution of the uniaxial component, $C_{u}$, in 40 and 10 nm thick films. All the AMR curves are recorded at 30 K under a 9 T magnetic field.}
	\label{fig:fig3}
\end{figure}

%%%%%%%%%%%%%%%%%%%%%%%%%%%%%%%%%%%%%%%%%%%%%%%%%%%%%%%%%%%%%%%%%%AMR%%%%%%%%%%%%%%%%%%%%%%%%%%%%%%%%%%%%%%%%%%%%%%%%%%%%%%%%%%%%%%%%%%%%%
\subsection{C. Magnetotransport Properties of \mpp\/ Thin Films}

After understanding the magnetic properties in \mpp\/ thin films, we proceeded to investigate their magnetotransport properties. We measured AMR in four types of Hall bar devices, aligned along different crystallographic directions viz. [110] - D1, [100] - D2, [$\Bar{1}$10] - D3, and [010] - D4, as shown in Fig.~\ref{fig:fig3}(a). To elucidate thickness dependence of AMR in \mpp\/ thin films, we present in Fig.\ref{fig:fig3}(b) the AMR data from Hall bar devices aligned along [110] in 40 and 10 nm thick films. While AMR in the 40 nm thick films show a two-fold symmetry, an additional four-fold component is clearly observed when film thickness is reduced to 10 nm. Fig.\ref{fig:fig3}(b) shows relative AMR, AMR($\phi$)$\%$ = $\frac{(\rho_{j}(\phi) - \rho_{avg})}{\rho_{avg}}$$\times$100 plotted as a function of the relative angle, $\phi$, between applied magnetic field and current directions, where $\rho_{avg}$ is the average value of longitudinal resistivity $\rho_{j}$ over a full 360\textdegree\/ rotation in the film plane. The observed AMR for different Hall bar devices, i.e. for current flowing along different crystallographic directions, shown in Fig.\ref{fig:fig3}(c-d), establish two important aspects. First, while the two-fold symmetric component is dependent on the relative angle between the magnetic field and current direction (non-crystalline term), the four-fold symmetric component is dependent on the relative angle between the magnetic field and [100] crystallographic direction of \mpp\/ (MgO [110] direction), and is independent of the current direction (crystalline term). Second, these two terms alone are not sufficient to explain the AMR observed for different current directions. For example, AMR amplitudes along \mpp\/[110] and \mpp\/[$\Bar{1}$10] are different for D1 and D3 devices, both in 10 and 40 nm thick films, which suggests the presence of a small, but finite uniaxial anisotropy along \mpp\/[110] in these thin films. The field dependence of the two-fold ($C_{2}$), four-fold ($C_{4}$), and uniaxial anisotropy ($C_{u}$) components are estimated by performing fits to the observed AMR with a phenomenological equation

\begin{align}
\begin{split}
    AMR\% = C_{2}cos(2\phi) + C_{4}cos(4\psi) + 
    C_{u}cos(2\chi)
\end{split}
\end{align}
  
where $\phi$, $\psi$, and $\chi$ represent angles between magnetic field vector $H$ and current direction $j$, [100], and [110] crystallographic directions, respectively (see Fig.\ref{fig:fig3}(a)). For the 40 nm thick films, $C_{2}$ shows a $H^{2}$ dependence (see Fig.~3(e)), but exhibits a saturation behavior at high fields for the 10 nm thick films (see Fig.~3(f)). In both the 10 and 40 nm thick films the magnitude of $C_{2}$ is much weaker in the D2/D4 devices compared to that in the D1/D3 devices. In addition, only in the 40 nm thick films the AMR in D1/D3 devices show a $\pi/2$ phase shift (see Fig.~3(c)). Both $C_{4}$, which is only present in the 10 nm thick films, and $C_{u}$ components increase with magnetic field strength, as shown in Figs.~3(g) and 3(h), respectively. However, $C_{u}$ is found to be much weaker compared to either $C_{2}$ or $C_{4}$, but of comparable magnitude in both 10 and 40 nm thick films. This indicates that the uniaxial AMR term in \mpp\/ thin films originates from the hetero-epitaxial interface and is likely due to anisotropic nucleation of growth at the MgO surface.

%%%%%%%%%%%%%%%%%%%%%%%%%%%%%%%%%%%%%%%%%%%%%%%%%%%%%%%%%%%%%%%%%%
%%%%Figure 4 %%%%%%%%%%%%%%%%%%%%%%%%%%%%%%%%%%%%%%%%%%%%%%%%%%%%%%%%%%%%%%%%%%
 \begin{figure} [h!]
	\centering
\includegraphics[width=1\linewidth]{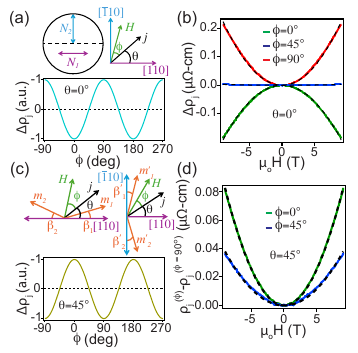}
	\caption{(a) Schematic of the two-domain model, as described in the text. The two-domain model predicts a two-fold AMR with a $\pi/2$ phase shift as observed for the D1/D3 devices in the experiment. (b) Evolution of $\Delta\rho_{j}=\rho_{j}(H)-\rho_{j}(H=0)$ with magnetic field for different $\phi$ values with $\theta=0^{\circ}$. The fits to eqn.~3, described in the main text, are shown with black dashed lines. (c) Canting of magnetic moments considered in the model described in the text. The spin canting model accurately predicts two-fold AMR for D2/D4 devices without any phase shift, in contrast to what is observed in the D1/D3 devices. (d) Evolution of $\Delta\rho_{j}^{(\phi)}=\rho_{j}^{(\phi)}-\rho_{j}^{(\phi=90^{\circ})}$ with magnetic field for different $\phi$ values with $\theta=45^{\circ}$. The fits to eqn.~5, described in the main text, are shown with black dashed lines.}
	\label{fig:fig4}
\end{figure}

%%%%%%%%%%%%%%%%%%%%%%%%%%%%%%%%%%%%%%%%%%%%%%%Magnetic Ground state%%%%%%%%%%%%%%%%%%%%%%%%%%%%%%%%%%%%%%%%%

%%%%%%%%%%%%%%%%%%%%%%%%%%%%%%%%%%%%%%%%%%%%%%%%%%%%%%%%%%%%%%%%%%
\subsubsection{Magnetoresistance in 40 nm thick films}

In order to understand the AMR and longitudinal magnetoresistance (MR) behavior in the 40 nm thick films, we construct a multi-domain model. Since \mpp\/ has a cubic crystal structure with $<$110$>$ as the easy axes\cite{suppl}, biaxial anisotropy dictates the presence of two energetically degenerate domains in absence of external magnetic field with the Neel vectors along [110]($N_{1}$) and [$\Bar{1}$10]($N_{2}$), as shown in Fig.~4(a). Application of magnetic field can induce either a coherent rotation of the Neel vectors in each of the domains or a spatial rearrangement of the domain configuration pushing the energetically favorable domain towards the unfavorable one. Typically, the latter is energetically favorable since it involves rotation of the spins only at the domain walls, which is of a significantly smaller volume fraction compared to the bulk of the domains\cite{fischer2018spin}. This rearrangement of the domains can give rise to MR as well as AMR. The longitudinal resistivity ($\rho_{j}$) in such a scenario can be written as 

\begin{align}
 \begin{split}
\rho_{j} = \rho_o + \rho_1 (\epsilon_1(N_1^{(j)})^2 + \epsilon_2(N_2^{(j)})^2)
 \end{split}   
\end{align}

with $\epsilon_1 + \epsilon_2$ = 1, where $\epsilon_i$ is the relative fraction of the domains with the Neel vector $N_{i}$ and $N_{i}^{j}$ is the projection of the Neel vector $N_{i}$ along the current direction $j$. The relative domain fractions $\epsilon_i$ under an applied field $H$ can be estimated by minimizing the total energy density ($E_{\text{total}}$) of the system (see Supplementary Information for details\cite{suppl}), allowing $\rho_{j}$ to be written as

\begin{align}
 \begin{split}
\rho_{j} = \rho_o + \frac{\rho_1}{2} \left[ 1 - \left( \frac{H_{\text{a}}}{2H_{\text{dest}}} + \frac{H^2}{4 H_{\text{ex}} H_{\text{dest}}} \cos 2(\theta + \phi) \right) \cos 2\theta \right]
 \end{split}   
\end{align}

where $H_{a}$, $H_{dest}$, and $H_{ex}$ are uniaxial anisotropy, destressing, and exchange fields, respectively. $\theta$ and $\phi$ are the angles between the current $j$ and crystallographic [110] directions, and magnetic field $H$, respectively (see Fig. 4(a)).

Therefore, for the D1($\theta$ = 0) and D3($\theta$ = $\pi/2$) devices, MR $\Delta\rho_{j}$ ($\Delta\rho_{j} = \rho_{j}(H) - \rho_{j}(0)$) is expected to exhibit $\Delta\rho_{j} \propto-H^{2}$,= 0, $\propto H^{2}$ dependence for $\phi = 0,\pi/4,\pi/2$, respectively, as is observed in our experiments, shown in Fig.~4(b). According to eqn.~3, AMR for both these devices is expected to exhibit -cos2$\phi$ = cos(2($\phi$+$\pi$/2)) dependence, explaining the $\pi/2$ phase shift observed in the experiments (see Fig.~3(b-c)). Moreover, the magnitude of the difference in resistivity between D1 and D3 devices for $\phi = n\pi/2, n\in\I$ (see Fig. 3(c)) is given by $\lvert\rho_{D1}(\phi = n\pi/2)-\rho_{D3}(\phi = n\pi/2)\rvert$ =$\rho_{1}H_{a}/2H_{dest}$. Assuming exchange coupling in \mpp\/ similar to that in MnPt, we estimate anisotropy field $H_{a}$ = 13.4 mT\cite{suppl,kang2023phonon}.

However, the present model fails to explain the AMR and MR behavior in the D2($\theta = -\pi/4$) and D4($\theta = \pi/4$) devices (see Fig.~3), even when accounting for small angular misalignments of the devices\cite{suppl}. The behavior of AMR and MR in these devices can be explained by considering the effects of spin canting. While this effect is not dominant in the D1 and D3 devices, it becomes significant in the D2 and D4 devices. This is because the MR and AMR due to domain reconfiguration under an applied field, which is the dominant effect, are zero in the D2 and D4 devices (see eqn.~3). Starting from the two energetically degenerate domains, as described in the earlier analysis, we now consider possible canting of the individual magnetic moments in both the domains, as shown in Fig. ~4(c). In such a scenario, the longitudinal resistance $\rho_{j} \propto M^{2}_{j}$\cite{nakagawa2023surface}, where $M_{j}$ is the net magnetic moment along $j$, which is identically zero in collinear antiferromagnets such as \mpp\/ in absence of spin canting. The canting angles for the magnetic moments $m_{1}$ and $m_{2}$ in the antiferromagnetic(AF) domain $N_{1}$, and for the moments $m^{'}_{1}$ and $m^{'}_{2}$ in the AF domain $N_{2}$ are $\beta_{1}$, $\beta_{2}$, $\beta^{'}_{1}$, and $\beta^{'}_{2}$, respectively (see Fig.~4(c)).  The net magnetic moment along the current direction, $j$, is obtained as 

\begin{align}
 \begin{split}
M_{j} =   \epsilon_{1}\times m [\cos(\theta - \beta_1)-  \cos(\theta + \beta_2)]+ \\
\epsilon_{2}\times m [\sin(\theta + \beta'_1)- \sin(\theta - \beta'_2)]\\
 \end{split}   
\end{align}

By minimizing the total energy (see Supplementary Information for details\cite{suppl}) we obtain relations for the canting angles, which allows us to write

\begin{align}
 \begin{split}
\rho_{j} \propto M_{j}^2 \sim \frac{H^2}{H_{ex}^2} \cos^2\phi\\
\Delta\rho_{j}^{(\phi)}=\rho_{j}^{(\phi)}-\rho_{j}^{(\phi=90^{\circ})} = \rho^{'}_{1}\frac{H^{2}}{2H^{2}_{ex}}(1+cos2\phi)\\
 \end{split}   
\end{align}

Therefore, we expect $\Delta\rho_{j}^{(\phi)}\propto H^{2}$ where $\Delta\rho_{j}^{(\phi)}$ = $\rho_{j}^{(\phi)}-\rho_{j}^{(\phi=90^{\circ})}$, as is observed in the experiments, shown in Fig.~4(d). Furthermore, the ratio of the coefficients of the $H^{2}$ dependence for $\phi = 0^{\circ}$ and $\phi = 45^{\circ}$ obtained from the fits in Fig.~4(d) is 2.1, similar to the value of 2 = $cos^{2}0^{\circ}/cos^{2}45^{\circ}$, expected from the theoretical model (see eqn.~5). Our model also explains the correct phase of the AMR for D2/D4 devices, shown in Fig.~3(a).

\begin{figure} [h!]
	\centering
\includegraphics[width=1\linewidth]{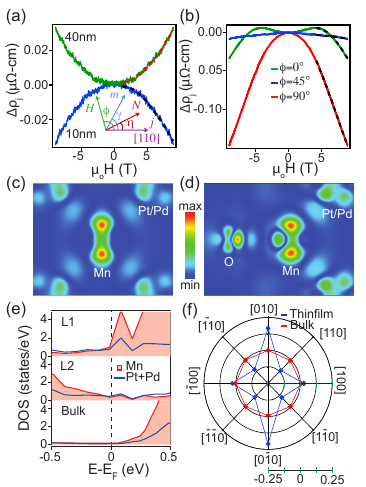}
	\caption{ (a) $\Delta \rho_{j}$ as a function of magnetic field applied along out-of-plane(001) direction recorded at 30 K for D1 devices of 40 nm and 10 nm thick \mpp\/ films. Inset shows a schematic that defines the angles described in eqn.~6, described in the main text. (b) In-plane MR for a 10 nm thick film in a D1 device for different magnetic field directions. In (a) and (b), fits to $Hln(H)$ and $H^{2}$ dependence are shown in black and red dashed lines, respectively. Charge density plots around the Mn atoms in \mpp\/ (c) in the bulk (d) at the \mpp\//MgO hetero-interface. Significant charge redistribution due to orbital overlap between O $p_{z}$ and Mn $d_{z^{2}-r^{2}}$ can be observed in (d). (e) Layer-resolved density of states of the \mpp\/ atomic layers. L1 and L2 correspond to the \mpp\/ layer at the interface and next to the interface, respectively. Partial DOS(PDOS) of Mn atoms is shown in red while the sum of PDOS of Pt and Pd atoms is shown in blue. DOS at E$_{F}$ is progressively reduced for \mpp\/ layers further away from the interface. (f) Change in DOS (CDOS) of \mpp\/ as a function of the direction of the Neel vector rotated within the (001) plane. CDOS is defined as the DOS (along a Neel vector) minus the average DOS. For the thin film calculation, Mn termination is considered (see text)} 
	\label{fig:fig5}
\end{figure}

Under the application of an out-of-plane magnetic field, 40 nm thick films exhibit a positive MR with a quadratic field dependence, irrespective of the in-plane current direction, shown in Fig.~\ref{fig:fig5}(a). This behavior is expected from Kohler's rule where MR is dominated by a single scattering time\cite{kohler1938magnetischen,chang2023electrical}. Kohler's scaling for 40 nm thick films is shown in the Supplementary Information (see Fig.~S10 in \cite{suppl}).

%%%%%%%%%%%%%%%%%%%%%%%%%%%%%%%%%%%%%%%%%%%%%%%%%%%%%%%%%%%%%%%%%%
%%%%Figure5 %%%%%%%%%%%%%%%%%%%%%%%%%%%%%%%%%%%%%%%%%%%%%%%%%%%%%%%%%%%%%%%%%%

\subsubsection{Magnetoresistance in 10 nm thick films}

Having understood the AMR and MR behavior in the 40 nm thick films, we now turn towards magnetotransport in the 10 nm thick films. In contrast to the 40 nm thick films, the 10 nm thick films do not exhibit Kohler's scaling\cite{suppl}, where the AMR and MR responses are strongly modified due to the presence of UM. For both out-of-plane and in-plane magnetic field configurations, shown in Figs.~5(a-b), negative MR is observed at higher fields exhibiting a $Hln(H)$ dependence indicative of a suppression of electron-magnon scattering.\cite{raquet2002electron} This behavior, which is typically observed in weak ferromagnetic metals, is driven by the presence of UM in the 10 nm thick films\cite{suppl}, which are easily polarizable under the application of magnetic field. To understand the behavior of the Neel vector on rotation of the magnetic field in the 10 nm thick films we write down the total energy of the system

\begin{align}
 \begin{split}
E_{total} = -MHcos(\phi-\gamma) + mH_{a}sin^{2}\eta - mH^{'}_{ex}cos(\gamma-\eta) 
 \end{split}   
\end{align}

where $M$, $m$, $H_{a}$, and $H^{'}_{ex}$ are magnetization due to UM, sub-lattce magnetization of the Neel order, antiferromagnetic anisotropy field, and exchange field corresponding to exchange interaction between UM and Neel order, respectively. $\phi$, $\gamma$, and $\eta$ are the angles between the crystallographic direction [110] and magnetic field, uncompensated moment, and Neel vector, respectively. Here, without loss of generality the current $j$ is taken along [110], which is also the antiferromagnetic easy axis. Moreover, we have neglected the anisotropy term for the uncompensated moments since they are easily polarizable by magnetic field. Our ab-initio calculations confirm that the easy axis of \mpp\/ lies along [110]\cite{suppl}, as is considered here. In addition, both from our calculations and magnetotransport data in 40 nm thick \mpp\/ films, we find that the anisotropy field $H_{a}$ in \mpp\/ is extremely small (see sections V and VIII in \cite{suppl}), consistent with the earlier results\cite{kumar2018crystal, umetsu2006magnetic, park2019strain}. It is obvious from eqn.~6 that in the limit $H^{'}_{ex}$ $\gg$ $H_{a}$, the energy is minimized for $\phi = \gamma$ and $\gamma = \eta$. Weak $H_{a}$ in \mpp\/ allows for the above limit to be easily achieved. As a result, we propose that the rotation of the magnetic field can drive a coherent rotation of the Neel vector in the 10 nm thick \mpp\/ films. This effect is due to the presence of UM and their interaction with the Neel vector. For coherent rotation of the spin-axis in the $a$-$b$ plane, AMR in \mpp\/ can be written as 

\begin{align}
 \begin{split}
AMR\% = C_{i}cos(2\phi) + C_{c}cos(4\psi) + \\
    C_{u}cos(2\chi) - C_{ic}cos(4\psi-2\phi)
 \end{split}   
\end{align}
which incorporates all the symmetry allowed terms\cite{ritzinger2023anisotropic,rushforth2007anisotropic,wang2020spin}, where $C_{i}$, $C_{c}$, $C_{u}$, and $C_{ic}$ represents two-fold non-crystalline, four-fold crystalline, and uniaxial crystalline terms, and an interaction term between non-crystalline and four-fold crystalline components, respectively. The angles have the same definition as in eqn.~1(see Fig.~3(a)). For D1/D3 devices $C_{2}$=($C_{i}$+$C_{ic}$), while for D2/D4 devices $C_{2}$=($C_{i}$-$C_{ic}$). The presence of $C_{ic}$ terms in the AMR explains the smaller magnitude of $C_{2}$ in D2/D4 devices compared to D1/D3 devices, as shown in Fig~3(f).

Since the crystalline component in AMR, as observed in the 10 nm thick \mpp\/ films, arises primarily due to modification in the equilibrium
relativistic electronic structure\cite{zeng2020intrinsic,dai2022fourfold}, we carried out ab-initio calculations to understand the effect of coherent rotation of Neel vector on electronic structure in \mpp\/ thin films. We performed calculations both for the bulk \mpp\/ (corresponding to 40 nm thick films) and thin film (slab) calculations, where the substrate was explicitly taken into account (corresponding to 10 nm thick films). Our calculations reveal a significant overlap between Mn $d_{z^{2} - r^{2}}$ and O $p_{z}$ orbitals at the interface that results in charge transfer onto the \mpp\/ atomic layers, which is observed in the charge density plots of Mn atoms near the interface(see Fig.\ref{fig:fig5}(c-d)). This leads to a large enhancement in the density of states (DOS) at the Fermi level. This effect is stronger for the \mpp\/ layers closer to the interface, which gradually reduces to zero in the bulk limit, as shown in Fig.\ref{fig:fig5}(e).  Rotation of the Neel vector in the $a$-$b$ plane leads to changes in the electronic structure resulting in a modulation of DOS near the Fermi level, which has a clear four-fold symmetry, as shown in Fig.\ref{fig:fig5}(f). The DOS has a peak near the Fermi energy for $<$100$>$ directions and a dip along $<$110$>$ directions. According to the Fermi golden rule, the electronic scattering rate is directly proportional to the DOS at the Fermi level. Hence, our calculation predicts a resistance maxima along $<$100$>$ and a minima along $<$110$>$ on rotation of the Neel vector, as observed in the experiments. Since the enhanced DOS contribution from the interface is only significant in the thinner films and reduces to zero in the bulk limit, the four-fold symmetric AMR is absent in the thicker 40 nm films(see Fig.\ref{fig:fig3}(c)). Similar results on interface induced enhancement of DOS in thinner films have also been obtained in earlier calculations of \la\/ antiferromagnets\cite{chang2021voltage}. The results shown in Fig.~5(f) was calculated with a Mn termination at the \mpp\//MgO interface (see Fig.~S9 and section VIII in \cite{suppl}), which is the more stable hetero-interface. However, similar results were also obtained assuming a Pt/Pd termination\cite{suppl}. Moreover, the charge transfer at the interface modifies the orbital occupancy of the Mn atoms(see Fig.\ref{fig:fig5}(d)) inducing additional moment\cite{suppl}. This results in a net UM in \mpp\/ thin films, which gets fully compensated in the bulk limit. 

In the 10 nm thick films, the uncompensated moments (UM) rotate in sync with both the magnetic field and the Neel vector, as discussed earlier. This coherent rotation contributes to the four-fold modulation of the DOS and the emergence of associated four-fold crystalline AMR. Disentangling the individual contributions of the UM and the Neel vector to the emergence of the four-fold symmetric AMR is therefore challenging. However, we note that since the four-fold crystalline AMR originates from the modifications in the electronic structure, it is more likely affected by the antiferromagnetic Neel order, which is the long-range magnetic order present in the bulk of these thin films.  Our ab-initio calculations demonstrate that the rotation of the Neel vector in collinear antiferromagnets can indeed lead to changes in the electronic structure, thereby inducing a four-fold symmetric crystalline AMR, as observed in the experiments. We also note that although hetero-epitaxial interface of MgO and \mpp\/ plays a major role in the observation of UM, we cannot rule out a contributory role of enhanced disorder in the 10 nm thick films in stabilizing UM, experimental evidence for which is provided in the Supplementary Information\cite{suppl}.

\section{Conclusion}

In summary, we have studied thickness dependence of magnetotransport in epitaxial thin films of a collinear antiferromagnet \mpp\/. Our findings reveal that the AMR in thicker films is influenced primarily by magnetic field-induced domain reconfiguration and spin canting of the sub-lattice magnetic moments. However, as the film thickness decreases, significant orbital overlap at the hetero-epitaxial interface between Mn $d$ and O $p$ orbitals along with finite disorder leads to the stabilization of UM, which induces a coherent rotation of the Neel vector under an applied magnetic field. Similar to ferromagnetic materials, the rotation of the Neel vector results in a strong modulation of the density of states at the Fermi level, which, in turn, gives rise to a four-fold symmetric crystalline component in AMR that is observable only in the thinner films. Our work has significant implications for designing and manipulating the Neel vector in antiferromagnets, which could play a crucial role as active elements in spintronic devices.

\section*{ACKNOWLEDGEMENTS}

We thank Devendra Buddhikot, Ruta Kulkarni, Debashis Mondal, and Ganesh Jangam for technical assistance. We acknowledge the Department of Science and Technology (DST), SERB grant SRG/2021/000414 and Department of Atomic Energy (DAE) of the Government of India (12-R$\&$D-TFR-5.10-0100) for support. We thank Dr. Sunil Ojha and Dr. G. R. Umapathy for Rutherford backscattering spectrometry (RBS) measurements on \mpp\/ thin films at Inter-University Accelerator Centre (IUAC), Delhi. We also acknowledge the use of the facilities at Industrial Research and Consultancy Centre (IRCC), IIT Bombay and National Nano Fabrication Centre (NNFC), IISc. 

S.Y., with assistance from S.K.G., performed thin-film growth. S.Y., N.K., and S.K.G, performed structural characterization of the thin films. B.C. and R.B. prepared thin-film lamellae. A.R., under the guidance of A.G., performed EDX characterization. S.Y. fabricated Hall bar devices. S.Y., with assistance from S.K.G. and D.P., performed magnetotransport measurements. M.V., under the guidance of A.K., performed ab-initio calculations. S.Y. and S.C. analyzed the data. S.Y. and S.C. wrote the manuscript. S.C. conceived the project and was responsible for its overall execution. All authors discussed the results and commented on the manuscript.

The authors declare no conflict of interest.

\section*{DATA AVAILABILITY}

The data supporting the findings of this article are openly available \cite{data}.

\section*{Supplemental Material}

Supplementary information contains 9 sections and 11 figures \cite{suppl, data}.

%\bibliography{References/bibMPP_updated}

\end{document}